\documentclass{ws-procs975x65} 
\usepackage{url}
\usepackage{amssymb}
\usepackage{amsmath}
\usepackage{hyperref}
\usepackage[usenames,dvipsnames]{color}
\begin{document}

\title{Superluminal Propagation and Acausality of Nonlinear Massive Gravity} 

\author{S. Deser}

\address{Lauritsen Lab, Caltech, Pasadena CA 91125, USA.\\
Physics Department, Brandeis University, Waltham, MA 02454, USA.\\
E-mail: deser@brandeis.edu}

\author{K. Izumi}
\address{Leung Center for Cosmology and Particle Astrophysics,\\ National Taiwan University, Taipei 10617, Taiwan.\\
E-mail: izumi@phys.ntu.edu.tw}

\author{Y. C. Ong}
\address{Leung Center for Cosmology and Particle Astrophysics,\\ National Taiwan University, Taipei 10617, Taiwan.\\
Graduate Institute of Astrophysics, National Taiwan University, Taipei 10617, Taiwan.\\
E-mail: ongyenchin@member.ams.org}

\author{A. Waldron}
\address{Department of Mathematics, University of California, Davis, CA 95616, USA.\\
E-mail: wally@math.ucdavis.edu}

\begin{abstract}
Massive gravity is an old idea: trading geometry for mass. 
Much effort has been expended on establishing a healthy model, culminating in the current ghost-free version. We summarize here our recent findings---that it is still untenable---because it  {likely admits \emph{local} acausalities: solutions with CTCs in a small neighborhood of any event.} 
\end{abstract}

\footnotetext[1]{We dedicate this contribution to Freeman Dyson on his 90th birthday. To Dyson's belief that the graviton---if exists---is unobservable, we add that it is in any case massless.}

\bodymatter

\section{Introduction: A Short History of Massive Gravity}

From a quantum field theory point of view, gravity is an interaction mediated by a spin-2 particle, the graviton~\cite{BD0}. A long-standing issue has been whether gravitons are truly massless like the gluon, or more like a light but massive neutrino (the latter has no gauge ``protection'', of course).

Massive gravity (mGR) can be traced back to Fierz and Pauli's (FP) 1939 formulation \cite{FP} of the free massive spin-2 field in flat space. They uniquely fixed it by requiring that it represents $2s+1=5$ (rather than the generic 6) excitations -- which then also guaranteed tachyon- and ghost-freedom. The FP action is the sum of the massless -- linearized GR -- kinetic and $m^2(h_{mn}^2 - c(h^m_{~m})^2)$ mass, terms and $c=1$ necessarily. But, as was realized much later \cite{vDVZ}, when stress-tensor sources $T_{mn}$ are included, their induced interactions take the form $\sim T_{mn}^2-a(m)(T^m_{~m})^2$, where  $a(m)=1/3$, while $a(0)=1/2$. This discrete discontinuity of course persists as $m \to 0$, leading to the well-known 25\% error in the massive model's light-bending prediction. 
The possibility remained that this disease was caused by taking the linear limit before the massless one~\cite{Vainshtein}.
However, this triggers a new -- and frightening -- obstacle: at nonlinear level there (generically) arises a massive, {\it ghost}, 6th degree of freedom beyond the 2$s$+1=5 of FP--the so-called Boulware-Deser (BD) ghost~\cite{BD}. Consequently, interest in mGR dwindled, until a---presumably consistent---BD-ghost-free extension was recently constructed \cite{dRGT}. These models involve, besides the dynamical metric $g$, a second, fixed, (fiducial) background metric $\bar{g}$; in tetrad formalism, $f^a_{~\mu}$ is the background (with inverse denoted by $\ell^\mu_{~a}$) and $e^a_{~\mu}$ the dynamical field. All index manipulations will be performed using the dynamical metric and tetrad, with Greek and Latin indices respectively representing world and local Lorentz coordinates.


The simplest example of nonlinear mGR has field equation 
\begin{equation}
G_{\mu\nu}(g) = \tau_{\mu\nu}(f,g) := \Lambda g_{\mu\nu} - m^2 \left(f_{\mu\nu} - g_{\mu\nu}f\right),
\end{equation}
where $f_{\mu\nu}=e^a_{~\mu}f^b_{~\nu}\eta_{ab}$ and $f=f^\mu_{~\mu}$. 
The parameter $m$ is the FP mass when the theory is linearized around a cosmological $\bar{\Lambda}$ background. The correct linearization requires that $\Lambda - \bar{\Lambda} + 3m^2 =0$. The tetrads obey the symmetry constraint $f_{[\mu}{}^m e_{\nu]m}=0$; its curl implies the integrability condition $f_{[\mu}{}^\sigma K_{\nu\rho]\sigma}=0$, where $K_{\mu}{}^m{}_n:=\omega_\mu(e){}^m{}_n-\omega(f)_\mu{}^m{}_n$ is the contortion and $\omega(e)_\mu{}^m{}_n$, $\omega(f)_\mu{}^m{}_n$ are the spin connections. 

Despite being ghost-free, subsequent investigations indicated that nonlinear mGR is still problematic. In particular, the characteristic equations in the eikonal limit were analyzed by two of the authors \cite{DW}. It was found that the model admits superluminal (second order) shock wave solutions, which ironically, is due to the very constraint that removes the BD ghost. Previously superluminal behavior was also uncovered in the model's St\"uckelberg sector and decoupling limit~\cite{Gr} as well as in a spherically symmetric analysis on Friedmann-Lema\^itre-Robertson-Walker~(FLRW) backgrounds~\cite{Chien-I}. One might think that since the graviton in this theory is \emph{massive}, it would automatically propagate \emph{slower} than light. However this is not necessarily the case. A simple counter-example is the nonlinear Proca field (massive photon) that also gives rise to a mode that propagates faster than light \cite{fT1}.

After the analysis, by the other two authors, of the characteristic matrix of the theory using PDE analysis \`a la Cauchy and Kovalevskaya \cite{IO},
the negative results of Refs.~\refcite{DW} and~\refcite{IO} were combined to show the massive theory not only gives rise to superluminal shock waves, but also local acausality \cite{DIOA} that can arise even in an infinitesimal neighborhood of a spacetime event, as summarized in the next section. 

\section{Superluminality vs. Acausality}

Superluminal shock waves can be found by studying discontinuities in the first derivative of fields across a hypersurface $\Sigma$, with normal $\xi$ chosen to be timelike; this is denoted by
\begin{equation}
[\partial_\alpha e_\mu{}^m]_\Sigma=\xi_\alpha{\cal E}_\mu{}^m ,\qquad 
[\partial_\alpha \omega_\mu{}^m{}_n]_\Sigma=\xi_\alpha \Omega_\mu{}^m{}_n.
\end{equation}
In particular we investigate the discontinuities in the scalar constraint (this equation does \emph{not} involve higher than first derivatives, and is responsible for removing the 6th, BD ghost, excitation)
\begin{equation}
0 = \frac{1}{m^2}\, \nabla_\rho \big(\ell^{\rho\nu}\nabla^{\mu}[{G}_{\mu\nu}-\tau_{\mu\nu}])+\frac{1}2\, g^{\mu\nu}
\,[{G}_{\mu\nu}-\tau_{\mu\nu}],
\end{equation}
and in the curl of the symmetry constraint. 
Denoting the contraction of~$\xi$ on an index of any tensor by an ``$o$'', so \emph{e.g.}, ~$\xi. V:=V_o$, 
upon carrying out a shock analysis of the theory's constraints, we obtain the characteristic matrix \cite{DIOA}
\begin{equation}
0=\begin{pmatrix}-\dfrac{3 m^2}{2}+\ell^{\mu}_o \big[{\bar R}_{\mu\nu}{}^{\nu}{}_o+K_{\mu\nu\rho} K^{\nu\rho}{}_o\big]& \frac 12\, \tilde K_j
\\[3mm]
[f\times K\ell]_i & f_{ij}-g_{ij} f^{(3)}
\end{pmatrix}\begin{pmatrix}{\cal F}_{oo}\\[3mm]\tilde\Omega^j\end{pmatrix}.
\end{equation}
Here ${\cal F}_{\mu\nu}:= {\cal E}_{\mu\rho}f_{\nu}{}^\rho$, $\tilde\Omega_i=\epsilon_{ijk}\Omega_{o}{}^{jk}$ and~$\tilde K_i=\epsilon_{ijk} K^{jk}{}_{o}$, where~$\epsilon_{ijk}:=\frac{1}{\sqrt{-g}}\, \xi^\mu \varepsilon_{\mu ijk}$, and $[f\times K\ell]_i:=2\, \epsilon_{ijk} f^{k\mu} K^{j}{}_{\nu\mu}\ell^\nu_o$ and~$f^{(3)}:=g^{ij}f_{ij}$. This gives a sufficient (but not necessary) condition for the \emph{field-dependent} determinant to vanish:
\begin{equation}
0=-\frac{3 m^2}{2}+\ell^{\mu}_o \big[{\bar R}_{\mu\nu}{}^{\nu}{}_o+K_{\mu\nu\rho} K^{\nu\rho}{}_o\big] -\frac 12\, \tilde K_i\ell^{ij}_{(3)}
[f\times K\ell]_j,
\end{equation}
where $\ell_{(3)}^{ij}:=\big(f_{ij}-g_{ij} f^{(3)}\big)^{-1}$. By choosing appropriate values of the fields, this determinant can be easily made to vanish, and so superluminal shocks are quite generic. An easy (and by no means even the only) demonstration of this fact is that for configurations such that $K_{\mu\nu o}=0$ in some region,
it follows that ${\cal F}_{oo}=0$ whence the condition for a vanishing determinant of the characteristic matrix reduces to, in an obvious matrix notation, $\left\{f, \Omega\right\}=0$. Thus,  the non-vanishing of sums of every pair $(f_1 + f_2, f_2 + f_3, f_1 + f_3)$ of eigenvalues of $f$ is sufficient and necessary for $\left\{f, \Omega\right\}=0$ to imply $\Omega=0$. However, the eigenvalues of $f$ are not necessarily positive, because of the difference in being ``spacelike'' with respect to the two metrics. Hence superluminal shocks can occur. Moreover, acausality is now very likely to arise since closed timelike curves can be locally embedded  into spacetime because spacelike surfaces can be characteristic ones as well. 

{To be more specific, specializing to a Minkowski background (say), consider the case when the temporal direction, together with one of the spatial coordinates (say the $3$-direction), of the fiducial and physical tetrads do not coincide. For example, if $f_{\mu\nu}= \text{diag}(1,1,1,-1)$,  then we have $f_1+f_3=0$ and $f_2+f_3=0$, {\it i.e.}, a constant time hypersurface is a characteristic. 
The action is invariant under simultaneous local Lorentz rotations of both physical and fiducial tetrads. 
Because of the interchange  of  the time- and $3$-direction some of these symmetries are ``spontaneously'' broken, while  $(1,2)$-rotation and $3$-boost symmetries are kept.
The $(2,3)$-rotation symmetry is broken (because of the flip of the $3$-direction, physical and fiducial tetrads rotate oppositely so 
physical and fiducial tetrads configurations change), but rotating by $\pi$, the new configurations thus generated become the same as the original ones. 
Combining with the $(1,2)$-rotation symmetry, the solution has still a parity symmetry with respect to the 3-direction. 
The time-constant characteristic hypersurface must support propagation in some direction
$(0,a,b,0)$ (say). 
Thanks to the  $(1,2)$-rotation symmetry  and parity symmetry with respect to $3$-direction, 
there must also be propagations in directions $(0,\pm a,b \cos\theta, b \sin \theta)$ where 
$\theta$ is an arbitrary constant.
Thus, we can construct closed timelike curves, for instance form a loop: $(0,a,b,0)$, $(0,a,0,b)$, 
$(0,-a,-b,0)$ and $(0,-a,0,-b)$.
Furthermore, since 
$3$-boosts do not change field configurations, and preserve $f_{\mu\nu}= \text{diag}(1,1,1,-1)$, the same method shows that closed timelike curves can be formed on the hypersurface generated by boosts.
Moreover, although a $1$-boost does not keep the original tetrad configurations, it will preserve the condition $f_2+f_3=0$,  and thus the boost-hypersurface is again characteristic. Because of $(1,2)$-rotational symmetry, the same holds if we consider a $2$-boost or mixture of $1$- and $2$-boosts.  By appropriate boosting, any spacelike hypersurface can be characteristic. 
We emphasize that we used flat background and field configurations $f_{\mu\nu}= \text{diag}(1,1,1,-1)$  purely for simplicity; it is not essential to our acausality.
}

Let us comment further on the difference between superluminality and acausality. 
In GR, we expect superluminal propagation to be also acausal, \emph{i.e.} we can construct closed timelike curves in the theory. 
However, in a theory that does not have coordinate invariance, superluminality does not always lead to acausality. For example, in Newtonian gravity, there is no upper speed limit, but there is no problem with acausality. In fact, even in a theory with diffeomorphism invariance and Lorentz invariance, one should remember that the mathematical structure of special relativity only requires \emph{existence} of an \emph{upper bound} on speed, and thus it could well be that the speed of light is very close to, but not really, the upper bound. Thus, superluminality is not always disastrous\cite{GBMV}. Nonlinear mGR is manifestly \emph{not} diffeomorphism invariant and the existence of superluminal shocks by itself is only an indication of \emph{possible} disaster, but existence of acausality means that the theory is \emph{definitively} bad. This scenario also happens in $f(T)$ gravity, which is not locally Lorentz invariant and thus can only be argued to be problematic via an acausality argument or the complete absence of predictability \cite{fT1, fT2}. Our investigations showed that nonlinear mGR does indeed admit acausality, which is not only much easier to construct than in GR, but is \emph{local} in nature. That is, unlike say, the G\"odel solution in GR, which in the neighborhood of any point of the closed timelike curve is still perfectly well-behaved (\emph{i.e.}, locally one always moves forward in time), the acausality in nonlinear mGR can be constructed in an infinitesimal regions of spacetime, and thus is much more serious. 
We mention that these problems are likely to persist in the bimetric extension of mGR, where the background field becomes dynamical~\cite{mc}.

\section{Conclusion: The State of Affairs}  

Ever since its conception, mGR has struggled to survive successive blows by adding successive epicycles (for a review, see Ref.~\refcite{Hinterbichler}). Our results demonstrate that its current incarnation is also untenable, due to the existence of not only superluminal shock waves, but also \emph{local} acausality. 
[The background metric is essentially an external field; such fields are well-known to generate acausalities in higher-spin theory contexts~\cite{velo}.]
Furthermore this acausality occurs quite generically, not only in the model's decoupling limit. This means that mGR \emph{cannot} be a UV-complete fundamental theory of gravity. One may argue that it can still be useful as an effective field theory. That is, it could still be well-behaved without acausality on some specific background, separated from the problematic ones by a putative potential barrier. However, since our acausality argument only depends on the rather weak condition that ${\cal F}_{oo}=0$, this situation is very unlikely. Furthermore, such a potential barrier cannot save the theory since it is no longer protected at the quantum level, especially since the natural scale of the theory is $\sqrt{M_{\text{planck}}m_\text{grav}}$, where $m_\text{grav}$, the graviton mass, is necessarily very small. Indeed, for $m_{\text{grav}} \sim H_0$, the value of such an energy scale is roughly $10^{-3}$eV, above which the effective field description is no longer applicable. [The absence of a supersymmetric (${\mathcal N}=1$) extension even of FP---simply because massive spin 3/2 has only $2s+1=4$ excitations---shows that SUSY's virtues are likely not available here either.]

The only possible way to remove the offending---ghost-removing, but also the cause of all our problems---scalar constraint would be to rely on the existence of a partially massless version of mGR; unfortunately, this last hope is also excluded, precisely at nonlinear level \cite{DW2}. It is gratifying that 
that GR and SU(3) YM are unique in both being exempt from the standard Higgs mass coupling mechanism, and in being ``isolated": not having viable ``neighboring" non-gauge models.

\section*{Acknowledgements}
We thank M. Porrati for requesting a CTC clarification, and W. Siegel for a SUSY query.
 S. Deser was supported in part by NSF PHY- 1266107 and DOE DE- FG02-164 92ER40701 grants. K. Izumi is supported by Taiwan National Science Council under Project No. NSC101-2811-M-002-103. Y. C. Ong was supported by the Taiwan Scholarship from Taiwan's Ministry of Education.


\begin{thebibliography}{100}

\bibitem{BD0} D.~Boulware  and S.~Deser Ann. Phys. {\bf 89} (1975) 193.

\bibitem{FP} M.~Fierz, W.~Pauli, 
Proc.~Roy.~Soc.~Lond.~A\textbf{173} (1939) 211.



\bibitem{vDVZ} H. van Dam, M. J. G. Veltman,  
Nucl. Phys. B \textbf{22} (1970) 397; V.~I.~Zakharov, 
JETP~Lett.~\textbf{12} (1970) 312. 

\bibitem{Vainshtein} A.~I.~Vainshtein, 
, Phys.~Lett.~B\textbf{39} (1972) 393; E.~Babichev, C.~Deffayet, R.~Ziour, 
Phys.~Rev.~D\textbf{82} (2010) 104008, \href{http://arxiv.org/abs/1007.4506}{[1007.4506v1[gr-qc]]}. 

\bibitem{BD} D.~Boulware, S.~Deser, 
Phys. Rev. D \textbf{6} (1972) 3368; 
Phys. Lett. B \textbf{40} (1972) 227.


\bibitem{dRGT} C.~de~Rham, G.~Gabadadze,  
Phys.~Rev.~D\textbf{82} (2010) 044020, \href{http://arxiv.org/abs/1007.0443}{[1007.0443v2  [hep-th]]}; C.~de~Rham, G.~Gabadadze, A.~J.~Tolley, 
Phys. Rev. Lett. \textbf{106} (2011) 231101, \href{http://arxiv.org/abs/1011.1232}{[1011.1232v2 [hep-th]]}.


\bibitem{DW} S.~Deser, A.~Waldron,  
Phys. Rev. Lett. \textbf{110} (2013) 111101, \href{http://arxiv.org/abs/1212.5835v3}{[1212.5835v3 [hep-th]]}.

\bibitem{Gr} A.~Gruzinov, 
\href{http://arxiv.org/abs/1106.3972}{[1106.3972 [hep-th]]}; C.~Burrage, C.~de Rham, L.~Heisenberg and A.~J.~Tolley, 
JCAP {\bf 1207} (2012) 004, \href{http://arxiv.org/abs/1111.5549}{[1111.5549 [hep-th]]}; P.~de Fromont, C.~de Rham, L.~Heisenberg and A.~Matas, 
\href{http://arxiv.org/abs/1303.0274}{[1303.0274 [hep-th]]}.


\bibitem{Chien-I} C.~-I.~Chiang, K.~Izumi, P.~Chen, 
JCAP \textbf{12} (2012) 025, \href{http://arxiv.org/abs/1208.1222}{[1208.1222v2 [hep-th]]}.


\bibitem{fT1} Y.~C.~Ong, K.~Izumi, J.~M.~Nester, P.~Chen, 
Phys. Rev. D \textbf{88}, 024019 (2013), \href{http://arxiv.org/abs/1303.0993}{[1303.0993 [gr-qc]]}.

\bibitem{IO} K.~Izumi and Y.~C.~Ong, 
Class. Quant. Grav. \textbf{30} (2013) 184008, \href{http://arxiv.org/abs/1304.0211}{[1304.0211 [hep-th]]}.
 
\bibitem{DIOA}  S.~Deser, K.~Izumi, Y.~C.~Ong, A.~Waldron, Phys. Lett. B \textbf{726} (2013) 544, 
\href{http://arxiv.org/abs/1306.5457}{1306.5457 [hep-th]]}.

\bibitem{GBMV} R. Geroch, 
\href{http://arxiv.org/abs/1005.1614}{[1005.1614 [gr-qc]]}; J-P. Bruneton, 
Phys. Rev. D 75 (2007) 085013, \href{http://arxiv.org/abs/gr-qc/0607055}{[gr-qc/0607055]}; N. Afshordi, D. J. H. Chung, G. Geshnizjani, 
Phys. Rev. D 75 (2007) 083513, \href{http://arxiv.org/abs/hep-th/0609150}{[hep-th/0609150]}; E. Babichev, V. Mukhanov, A. Vikman, JHEP 0802 (2008) 101, \href{http://arxiv.org/abs/0708.0561}{0708.0561 [hep-th]}.

\bibitem{fT2} K.~Izumi, J-A.~Gu, Y.~C.~Ong, 
\href{http://arxiv.org/abs/1309.6461}{[1309.6461 [gr-qc]]}.

\bibitem{mc} S.~Deser, M.~ Sandora and A.~Waldron, Phys Rev. D \textbf{88} (Rapid Communication) (2013) 081501,
\href{http://arxiv.org/abs/1306.0647}{[1306.0647 [hep-th]]}.

\bibitem{Hinterbichler} K.~Hinterbichler, 
Rev.~Mod.~Phys.~\textbf{84} (2012) 671, \href{http://arxiv.org/abs/1105.3735}{[1105.3735v2 [hep-th]]}.



\bibitem{velo}
K.~Johnson and E.C.G.~Sudarshan, Ann. Phys. {\bf 13} (1961) 126; 
G.~Velo and D. Zwanziger, Phys.Rev. {\bf 186} (1969) 1337;
M.~Kobayashi and A.~Shamaly, Phys.\ Rev.\ D {\bf 17}, (1978) 2179;
Prog.\ Theor.\ Phys.\  {\bf 61} (1979) 656 (1979);
S.~Deser and A.~Waldron, Nucl.\ Phys.\ B {\bf 631} (2002) 369,
 \href{http://arxiv.org/abs/hep-th/0112182}{[hep-th/0112182]}. 
 
\bibitem{DW2} 
  S.~Deser, M.~Sandora, A.~Waldron, 
   Phys. Rev. D \textbf{87} (2013)  101501(R) 
  \href{http://arxiv.org/abs/1301.5621}{[1301.5621 [hep-th]]}. 
 
 \end{thebibliography}
\end{document}